\documentclass[conference]{IEEEtran}
\IEEEoverridecommandlockouts
\usepackage[ruled,lined]{algorithm2e}
\usepackage{setspace}
\usepackage{float}
\usepackage{cite}
\usepackage{graphicx}
\usepackage{epstopdf}
\usepackage{graphicx,epsf,epic,eepic,psfig,latexcad}
\usepackage{amsmath,amssymb,amsfonts}
\usepackage{algorithmic}
\usepackage{textcomp}
\usepackage{xcolor}
\newcommand\numeq[1]%
{\stackrel{\scriptscriptstyle(\mkern-1.5mu#1\mkern-1.5mu)}{=}}

\def\BibTeX{{\rm B\kern-.05em{\sc i\kern-.025em b}\kern-.08em
		T\kern-.1667em\lower.7ex\hbox{E}\kern-.125emX}}
\begin{document}

\title{Approximate Weight Distribution of Polarization-Adjusted Convolutional (PAC) Codes}
\author{\IEEEauthorblockN{Sadra Seyedmasoumian and Tolga M. Duman}
\IEEEauthorblockA{\textit{Dept. of Electrical and Electronics Engineering,}
\textit{Bilkent University}, Ankara, Turkey \\
\{sadra,duman\}@ee.bilkent.edu.tr}}
\maketitle
\begin{abstract}
Polarization-adjusted convolutional (PAC) codes combine the polar and convolutional transformations to enhance the distance properties of polar codes. They offer a performance very close to the finite length information-theoretic bounds for short block lengths. In this paper, we develop a method of computing the weight distribution of PAC codes in an approximate form by employing a probabilistic technique. We demonstrate that the results well match the exact weight distributions for small codes that can be computed using a brute-force algorithm. We also present a way employing the results (along with the union bound on the code performance) to design specific PAC codes, more precisely, to determine suitable rate profiles via simulated annealing. Numerical examples illustrate that the PAC codes with the designed rate profiles offer superior performance.

\end{abstract}


\section{Introduction}
Polar codes \cite{Arikan2009} are the first deterministic class of codes that achieve binary-input symmetric channel capacity; however, due to their inferior distance properties, for practical block-lengths, their error-correcting performance is limited, and modifications are required in both the encoder and the decoder structures. Among different approaches, one widely used technique is the concatenation of polar codes with an outer CRC encoder \cite{Niu2012a}. With a suitably designed successive cancellation list decoder, the CRC-aided polar codes show significant improvements compared to the standard polar codes, and they find applications in 5G new radio (NR) systems. Another modification is a unique concatenation of polar codes with an outer convolutional code, i.e., the polarization-adjusted convolutional (PAC) codes \cite{Arkan2019}. 

PAC codes can be thought as polar codes with dynamically frozen bits that transfer the load of the error correction towards the convolutional decoder's side \cite{yao2021list}. There are two components in the construction of PAC codes: selection of the convolutional encoder and design of the rate profile, which is the process of selecting the frozen bit-positions from a set of $N$ potential indices where $N$ is the codeword length. The specific rate profile of a PAC code significantly impacts the decoder's error-correction performance and computational complexity. Polar and Reed-Muller (RM) rate profiles are two commonly used PAC code rate profiles. At the decoder side, the main operation is sequential decoding, in particular, Fano decoder as employed in \cite{Arkan2019}. Stack algorithm, which has a lower computational complexity for low signal-to-noise ratios (SNRs), can also be used \cite{mozammel2021hardware}. Other approaches include the successive cancellation list decoding algorithms as introduced in \cite{yao2021list}, \cite{Rowshan2020} and \cite{zhu2020fast}. Simulation results in \cite{Arkan2019} show that the PAC codes with sequential decoding are capable of providing better codeword error rates than 5G-standardized polar codes. The results depict that, for some cases, the PAC code error rate performance achieves the finite length information-theoretical limits with a Fano decoder and Reed-Muller score construction \cite{Arkan2019, Moradi2020}. We also note that several recent papers \cite{Moradi2021a,mishra2021modified,Sun2021} propose different approaches for rate-profile design in order to improve the error correction capability of PAC codes. 

Our focus is the weight distribution of PAC codes. Under maximum likelihood (ML) decoding, the weight enumerating function (WEF) can provide an accurate performance assessment, particularly, for high signal-to-noise ratios. We focus on approximating the WEF of PAC codes using an extension of the ideas in \cite{Valipour2013} and \cite{Zhang2017}. That is, we split the original PAC codewords into two sets of PAC codewords with half the length and continue this procedure until a predefined threshold is reached. At that point, we compute the weight distribution of the small codes exhaustively, and employ a recursive procedure to reconstruct the WEF of the original code in a probabilistic fashion leading to an approximation. With the obtained approximate WEF of PAC codes, we calculate (approximate) union bounds on the code performance (attainable under ML decoding). We also employ a discrete optimization method to design rate profiles that improve the error correction performance of PAC codes by using the derived union bound at a fixed SNR as a cost metric. Examples illustrate that this design procedure is quite robust; and, while it is based on ML decoding, it offers excellent performance even under practical decoding algorithms.

The rest of the paper is organized as follows. The structure of the PAC codes is introduced, and notation is established in Section \ref{sec2}. Section \ref{sec3} describes the proposed approximation method for computing the WEF of PAC codes. In Section \ref{sec4}, the PAC code design based on the derived approximate WEF is presented. Simulations and numerical results are given in Section \ref{sec5}, and finally, the paper is concluded in Section \ref{conc}.

\section{Preliminaries}
\label{sec2}
Consider a PAC code represented by $ \left(N,K,\mathcal{A}, \boldsymbol{T}_N \right) $ where $N$ and $K$ are codeword and message lengths, respectively, $\mathcal{A}$ is the set of information (non-frozen) bit-positions and $ \boldsymbol{T}_N \in \mathbb{R}^{N \times N}$ is an upper-triangular Toeplitz matrix constructed with the convolutional code generator $\boldsymbol{g}= \left\{g_1,g_2,...g_\ell \right\}$, where it is assumed that $g_0$, $g_\ell \neq 0$ (see \cite{Arkan2019}).

\begin{figure}
    \centering
    \includegraphics[width=3.5in, height=1.3in]{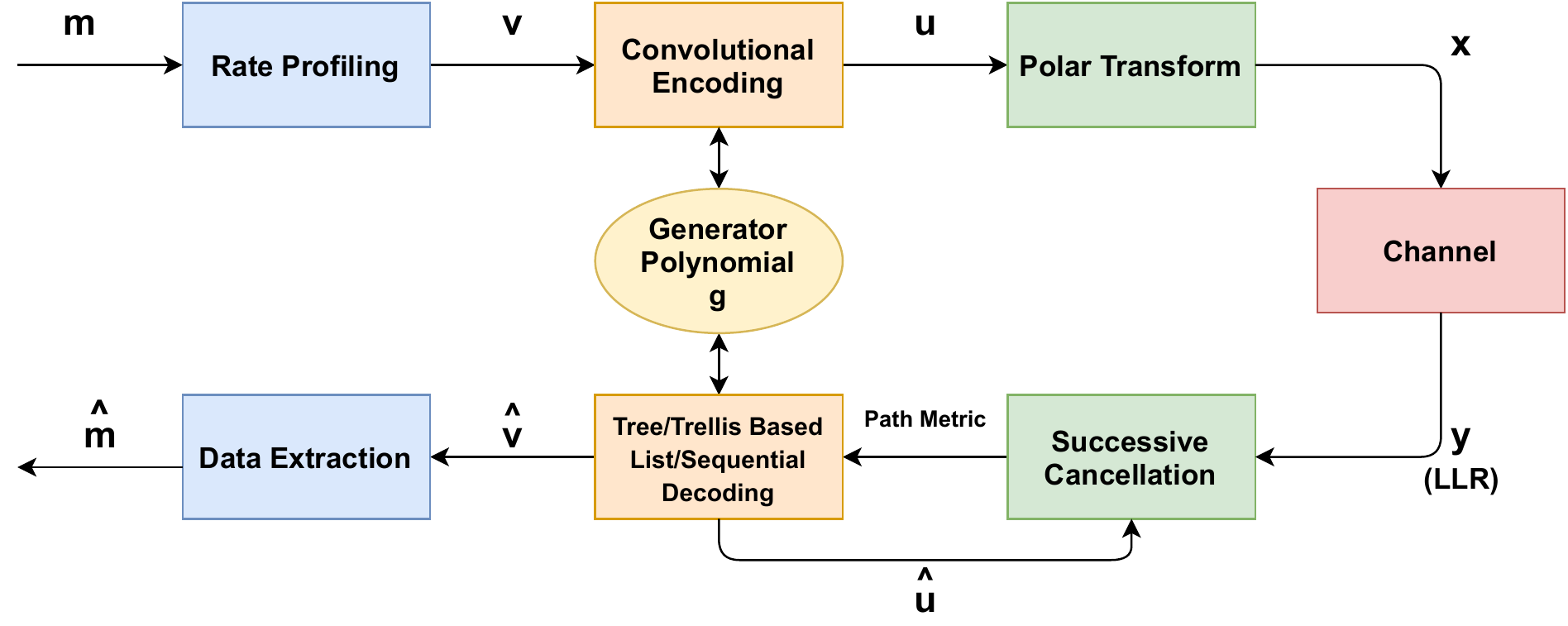}
    \caption{The block diagram of PAC code}
    \label{blockdiag}
\end{figure}

The encoding and decoding steps of PAC codes are depicted in Fig. \ref{blockdiag}. The encoder consists of three phases. In the first phase, $K$ information bits $\boldsymbol{m}=[m_0,m_1,...,m_{K-1}]$ are mapped to $K$ information bit-locations in the input sequence $\boldsymbol{v}$ of length $N$ using the information index set $\mathcal{A}$ where $\boldsymbol{v}_{\mathcal{A}}=\boldsymbol{m}$, and the cardinality of $\mathcal{A}$ equals $K$ ($\mathcal{|A|}=K$). The remaining $N-K$ bit-locations ($\mathcal{A}^c$) are called the frozen bit-locations, and $\boldsymbol{v}_{\mathcal{A}^c}=0$ is selected. This phase is the \emph{rate-profiling} step. The rate profile of a PAC code has a tremendous effect on its error-correcting performance. As an effective approach, Reed-Muller rate profiling is employed in \cite{Arkan2019}, where the bit-locations with higher RM scores are used as the information set. We note that for some parameters, we need to pick only a portion of the bit-locations with identical RM scores, which makes identification of the optimal frozen (or, information) bit-locations challenging. 

The second phase in the encoding process is the one-to-one convolutional encoding step. We denote the output of the convolutional encoder by $\boldsymbol{u}$. The $i$th convolutional bit is obtained by $u_i = \sum_{j = 0}^\ell g_j v_{i-j}$, where $v_{i-j} = 0$ for $i < j$, i.e., $\boldsymbol{u}=\boldsymbol{v} \boldsymbol{T}_N$.
In the third phase of the PAC encoder, the output of the convolutional encoder is fed to a polar encoder. That is, the codeword is obtained as $\boldsymbol{x}=\boldsymbol{u}\boldsymbol{G}_N$, where
\begin{equation}
    \boldsymbol{G}_N=\begin{bmatrix}
		1 & 0\\
		1 & 1
		\end{bmatrix}^{\otimes log_2(N)},
\end{equation}
where $\otimes$ represents the Kronecker product. 

The decoding algorithm for a PAC code can be implemented as a tree search algorithm such as Fano decoder or a list decoder \cite{Arkan2019, yao2021list,Rowshan2020,zhu2020fast}.

\section{Approximate Weight Distribution of PAC codes}
\label{sec3}
Computation of weight distribution of a linear block code is extremely difficult unless the code exhibits sufficient symmetry. This difficulty persists for polar or PAC codes as well, in particular, for practical code parameters which exclude the possibility of using an exhaustive search. There are several papers including \cite{Liu2014, polyanskaya2020weight} that address the weight distribution computation of polar codes. In addition, a probabilistic method for computing the weight distribution of polar codes is introduced in \cite{Valipour2013}, and it is later enhanced in \cite{Zhang2017}. A computation method for finding the WEF of pre-transformed polar codes (including PAC codes) is given in \cite{li2021weight}; however, this work assumes that the elements of the upper-triangular matrix are independent and identically distributed (i.i.d) \emph{Bernoulli}$(\frac{1}{2})$ random variables. In \cite{Yao2021}, a method to calculate the WEF of a polar code is provided, and some clues are given on how to extend it for the WEFs of polar codes with dynamically frozen bits, such as PAC codes. On the other hand, it does not seem feasible to employ the approach when $K$ is large. Here, with the main objective of providing a tractable WEF computation approach, we aim at developing an approximate probabilistic WEF computation method for PAC codes inspired by \cite{Valipour2013, Zhang2017}.

Define the weight distribution of a PAC code as
    \begin{equation}
        A(Z)=\sum_d A_d Z^d,
    \end{equation}
    where $A_d$ is is the number of codewords with weight $d$. When the information length is $K$, the total number of codewords is $2^K$. Let $P^{(d)}$ be the the probability that a random selected codeword has weight $d$. Clearly, $P^{(d)}=A_d/{2^K}$, i.e.,
    \begin{equation}
        A(Z)=\sum_d P^{(d)}2^K Z^d.
    \end{equation} 
\\
We split the codeword $\boldsymbol{x}$ into two parts, i.e., $\boldsymbol{x}=\{x_1^{N/2},x_{N/2+1}^{N}\}$ where\footnote{Throughout this paper, we use ${x}_a^b$ to represent $(x_a, x_{a+1},...,x_b)$, a subvector of $\boldsymbol{x}$.} $x_{N/2+1}^{N}=u_{N/2+1}^{N} \boldsymbol{G}_{N/2}$, and $x_1^{N/2}=(u_{N/2+1}^{N} \oplus u_1^{N/2}) \boldsymbol{G}_{N/2}$.
       Furthermore, we can write $\boldsymbol{T}_N$ as
       \begin{equation}
 \boldsymbol{T}_{N}= \begin{bmatrix}
           \boldsymbol{T}_{N/2} & \boldsymbol{S}_{N/2} \\
           \boldsymbol{0} & \boldsymbol{T}_{N/2}
       \end{bmatrix}
       \end{equation}
       where ${\boldsymbol{T}_{N/2}}$ is the convolutional encoding matrix of a PAC code with length $N/2$, and 
       \begin{equation}
       \label{SN2}
           \boldsymbol{S}_{N/2}=\begin{pmatrix}
               0 & \cdots & & & \cdots & & 0  \\
               \vdots & & \ddots & & & \ddots\\
               0& \cdots & & \cdots & & \cdots & 0 \\
               g_\ell & 0 & \cdots &  & \cdots & & 0\\
               g_{\ell-1} & g_\ell & 0 & \cdots & & \cdots & 0 \\
               \vdots & & \ddots & \ddots & & & \vdots \\
               g_1 & g_2 & \cdots & g_\ell & 0 & \cdots & 0
           \end{pmatrix}.
       \end{equation}
       The polar transformation matrix also can be wrriten as
       \begin{equation}
           \boldsymbol{G}_N= \begin{bmatrix}
               \boldsymbol{G}_{N/2} & \boldsymbol{0}\\
                \boldsymbol{G}_{N/2} & \boldsymbol{G}_{N/2}
           \end{bmatrix}.
       \end{equation}
       Therefore,
       \begin{equation}
           \boldsymbol{x}=\boldsymbol{v}\begin{bmatrix}
           \boldsymbol{T}_{N/2} & \boldsymbol{S}_{N/2} \\
           \boldsymbol{0} & \boldsymbol{T}_{N/2}
       \end{bmatrix}
         \begin{bmatrix}
               \boldsymbol{G}_{N/2} & \boldsymbol{0}\\
                \boldsymbol{G}_{N/2} & \boldsymbol{G}_{N/2}
           \end{bmatrix} 
       \end{equation}
       Thus, 
       \begin{multline}
           x_1^{N/2}=v_1^{N/2} \boldsymbol{T}_{N/2} \boldsymbol{G}_{N/2} \oplus v_{N/2+1}^{N} \boldsymbol{T}_{N/2} \boldsymbol{G}_{N/2} \\ \oplus v_1^{N/2} \boldsymbol{S}_{N/2} \boldsymbol{G}_{N/2}, 
       \end{multline}
       and
       \begin{equation}
         x_{N/2+1}^{N} = v_1^{N/2} \boldsymbol{S}_{N/2} \boldsymbol{G}_{N/2} \oplus v_{N/2+1}^{N} \boldsymbol{T}_{N/2} \boldsymbol{G}_{N/2}.
       \end{equation}
         We note that, $v_{N/2+1}^{N} \boldsymbol{T}_{N/2} \boldsymbol{G}_{N/2}$, and $v_1^{N/2} \boldsymbol{T}_{N/2} \boldsymbol{G}_{N/2}$ are PAC codes with $(N/2,|\mathcal{A}_2|,\mathcal{A}_2,\boldsymbol{T}_{N/2})$ and $(N/2,|\mathcal{A}_1|,\mathcal{A}_1,\boldsymbol{T}_{N/2})$, corresponding to the input sequences $v_{N/2+1}^{N}$ and $v_1^{N/2}$, respectively, and $\mathcal{A}_1=\left\{i \in \mathcal{A} | i \leq \frac{N}{2}\right\}$, and $\mathcal{A}_2=\left\{i \in \mathcal{A} | i > \frac{N}{2}\right\}-\frac{N}{2}$.
         If we represent $v_1^{N/2} \boldsymbol{T}_{N/2} \boldsymbol{G}_{N/2}$, $v_{N/2+1}^{N} \boldsymbol{T}_{N/2} \boldsymbol{G}_{N/2}$ and $v_1^{N/2} \boldsymbol{S}_{N/2} \boldsymbol{G}_{N/2}$ as $\boldsymbol{x}^{(1)}$, $\boldsymbol{x}^{(2)}$ and $\boldsymbol{s}$, respectively, the Hamming weight of $x_1^N$ is given by
         \begin{equation}
         \label{hamming}
    h \left ( x_1^N\right )= h\left( \boldsymbol{x}^{(1)} \oplus \boldsymbol{x}^{(2)} \oplus \boldsymbol{s} \right) +  h\left(\boldsymbol{x}^{(2)} \oplus \boldsymbol{s}\right),
\end{equation}
\noindent where $h(\boldsymbol{x})$ denotes the Hamming weight of sequence $\boldsymbol{x}$.\\
If we ignore the effect of the vector $\boldsymbol{s}$ (i.e., assume $\boldsymbol{s}=\boldsymbol{0}$), Eq. (\ref{hamming}) is reduced to
 \begin{equation}
         \label{hamming2}
    h \left ( x_1^N\right )= h\left( \boldsymbol{x}^{(1)} \oplus \boldsymbol{x}^{(2)} \right) +  h\left(\boldsymbol{x}^{(2)}\right)
\end{equation}
That is, in each step, we effectively split the PAC code into two PAC codes of half the length with information sets of $\mathcal{A}_1$ and $\mathcal{A}_2$, respectively. For a given $\boldsymbol{x}^{(2)}$ vector, we model the $\boldsymbol{x}^{(1)}$ vector randomly, and write an approximate weight distribution in a probabilistic fashion. We compute the approximate weight distribution of the code as

    \begin{multline}
    \label{recursive}
    P^{(d)}[\mathcal{A},N,\boldsymbol{T}_N] = \sum_{d_1} P\left(h(\boldsymbol{x}^{(2)})=d_2|\mathcal{A}_2\right)
      \\ \times P\left(h(\boldsymbol{x}^{(1)} \oplus \boldsymbol{x}^{(2)})=d-d_2|\mathcal{A}_1 , h(\boldsymbol{x}^{(2)})=d_2\right) 
      \\=\sum_{d_2} \sum_{d_1} P^{(d_2)} [\mathcal{A}_2,N/2,\boldsymbol{T}_{N/2}]  \\ \times P^{(d_1)} [\mathcal{A}_1,N/2,\boldsymbol{T}_{N/2}] \times f_N(t,d_1,d_2)
    \end{multline}
   with $t=d-d_2$, and
   \begin{multline}
       f_N(t,d_1,d_2)=P \Big( h\left( \boldsymbol{x}^{(1)} \oplus \boldsymbol{x}^{(2)} \right) =t | h(\boldsymbol{x}^{(1)})=d_1,  \\  h(\boldsymbol{x}^{(2)})=d_2 \Big).
       \label{xx}
   \end{multline}
   In other words, the weight distribution of a PAC code with length $N$ can be recursively obtained through that of a PAC code with length $N/2$ until a threshold value of $N_{th}$. We can then evaluate the remaining weight distributions using exhaustive search to enhance the accuracy of the approximation. We can use a similar function to $f_N$ defined in \cite{Valipour2013,Zhang2017}, given by $f_N(t,d_1,d_2)= \binom{d_1}{r} \binom{N/2-d_1}{d_2-r}/\binom{N/2}{d_2}$ for $t\geq|d_1-d_2|$ or $t \leq min\{d_1+d_2,N-d_1-d_2\}$, and $0$ otherwise, where $r=(d_0+d_1-t)/2$ is the number of positions at which the elements in $\boldsymbol{x}^{(1)}$ and $\boldsymbol{x}^{(2)}$ are both equal to 1. 
   
To increase the accuracy of the approximation, we can consider the actual value of the $\boldsymbol{s}$. We divide the whole sequence into two parts with information index sets of $\mathcal{A}_1$ and $\mathcal{A}_2$ until we reach a threshold codeword length $N_{th}$. After that we compute the weight distribution of each part in an exact manner by taking the matrix $\boldsymbol{S}_{N_{th}/2}$ into account, exhaustively. As depicted in Eq. (\ref{SN2}), $\boldsymbol{S}_{N/2}$ is a sparse matrix and has an effect only on $\ell$ bits, i.e., on $v_{\frac{N}{2}-\ell+1}^{\frac{N}{2}}$. For the rest of the section, we represent $v_{\frac{N_{th}}{2}-\ell+1}^{\frac{N_{th}}{2}}$ by $\boldsymbol{v}^{(\boldsymbol{s})}$. For calculating $h(\boldsymbol{x}^{(2)} \oplus \boldsymbol{s})$, we fix  $\boldsymbol{v}^{(\boldsymbol{s})}$, and find the weight distribution of $\boldsymbol{x}^{(2)}$ with codeword length $N_{th}/2$ and the set of information indices $\mathcal{A}_2$. We will have a list of $2^{|\mathcal{A}_s|}$ weight distributions for $\boldsymbol{x}^{(2)} \oplus \boldsymbol{s}$, where $\mathcal{A}_s=\left\{i \in \mathcal{A}_1| i > N_{th}/2-\ell\right\}$, which are the information bit locations related to $\boldsymbol{v}^{(\boldsymbol{s})}$ . On the other hand, with fixed $\boldsymbol{v}^{(\boldsymbol{s})}$ we estimate the weight distribution of $\boldsymbol{x}^{(0)}$, which is the set of indices that are not produced by $\boldsymbol{v}^{(\boldsymbol{s})}$ and has information bit sets of $\mathcal{A}_0=\left\{i \in \mathcal{A}_1| i \leq N_{th}/2-\ell\right\}$. 

Let $P^{(d)}[\mathcal{A},N_{th}]$ denote the probability that a PAC code-word with length $N_{th}$, information set $\mathcal{A}$ and polynomial $\boldsymbol{g}$ that generates $\boldsymbol{T}_{N_{th}}$ has weight $d$. We can calculate $P^{(d)}[\mathcal{A},N_{th}]$ by 
    \begin{multline}
    \label{partly}
       P^{(d)}[\mathcal{A},N_{th}]=\sum_{\boldsymbol{v}^{(\boldsymbol{s})}} \sum_{d_1} \sum_{d_0}P\left( h(\boldsymbol{x}^{(0)})=d_0|\mathcal{A}_0 ,\mathcal{A}_s\right) \\ \times P\left(h(\boldsymbol{x}^{(2)}\oplus \boldsymbol{s})=d_1|\mathcal{A}_2,\mathcal{A}_s\right) f_{N_{th}}(t,d_1,d_0),
    \end{multline}
        with $t=d-d_1$, since $P\Big( h\left( \boldsymbol{x}^{(1)} \oplus \boldsymbol{x}^{(2)}+ \boldsymbol{s} \right)=t| h\left(\boldsymbol{x}^{(2)}+\boldsymbol{s}\right)=d_1, h\left(\boldsymbol{x}^{(0)}\right)=d_0\Big)$ is simply $f_{N_{th}}(t,d_1,d_0)$.
 
    After finding the weight distribution of each part by (\ref{partly}), we use (\ref{recursive}) to find the weight distribution of PAC code with length $N$, recursively.
  
\section{Rate Profiling Based on the Approximate WEF}
\label{sec4}
Using the approximate WEF for a PAC code developed in the previous section, we can obtain an approximate union bound on the codeword error probability over an AWGN channel with BPSK modulation as
\begin{equation}
    P_e \leq \sum_{d=1}^{N-1} A_d P(d)
\end{equation}
where $P(d)$ is the pairwise error probability, also denoted by $P(d)=P(\boldsymbol{c} \to \boldsymbol{\hat{c}})$, between two codewords ($\boldsymbol{c}$ and $\boldsymbol{\hat{c}}$) with Hamming distance of $d$, given by
\begin{equation}
\label{union}
   P(\boldsymbol{c} \to \boldsymbol{\hat{c}}) =Q\left(\sqrt{\cfrac{\gamma_s d_E^2(\boldsymbol{c},\boldsymbol{\hat{c}})}{2}}\right).
\end{equation}
Here $\gamma_s$ is the SNR (i.e., $\gamma_s=\frac{Es}{N_0}$), and $d_E^2(\boldsymbol{c},\boldsymbol{\hat{c}})=4d$, assuming BPSK modulation with unit symbol energy.
\begin{algorithm}[t!]
\label{algorithm1}
\caption{Simulated Annealing}
\hspace*{\algorithmicindent} \textbf{Input} $N,\,K,\,\boldsymbol{g},\,\text{Starting}\,\mathcal{A},\, T_{max}, T_{min}, \, a $ \\
 \hspace*{\algorithmicindent} \textbf{Output} best\_inf 
 \begin{algorithmic} 
\STATE best\_E=1; 
 i=1;
T=Tmax;
\WHILE{$T>T_{min}$}
\STATE $E_c \leftarrow cost(N,K,g,\mathcal{A})$
\STATE $\mathcal{A}\_next$ $\leftarrow$ Perturb a single bit location
\STATE $E_n \leftarrow cost(N,K,g,\mathcal{A}\_next)$;
 $\Delta=E_n-E_c$
\IF{$\Delta<0$}
\STATE $i \leftarrow i+1$;
Info $\leftarrow$ $\mathcal{A}\_next$;
       
        \IF{$ En< best\_E$}
       \STATE best\_inf $ \leftarrow $ Info;
         best\_E  $\leftarrow$ $E_n$
        \ENDIF
       \STATE T $\leftarrow T_{max} \times a^{i}$ \label{op11}
\ELSIF{$e^{(-\Delta/T)}>rand(0,1)$}
\STATE $i \leftarrow i+1$;
Info $\leftarrow $ $\mathcal{A}\_next$;
\STATE T $\leftarrow T_{max} \times a^{i}$ \label{op12}
\ENDIF
\ENDWHILE
\end{algorithmic}
\end{algorithm}

 \vspace{0.3cm}
\noindent\emph{Rate Profiling via Simulated Annealing}

\noindent The number of possible rate-profiles that can be generated for a code with a codeword length of $N$ and a rate of $R$ is $\binom{N}{RN}$, which is a very large number, and running through every possible codeword with a brute-force algorithm is impossible. Equipped with a simple and efficient method to calculate the union bound of a given PAC code, we can use the bound's value at a fixed SNR as the cost, and perform metric optimization among different rate profiles. Specifically, we can employ Simulated Annealing (SA) to explore a set of different rate profiles and discover the one with the lowest cost metric, i.e., the best error correction performance under maximum likelihood (ML) decoding.

Simulated annealing is a probabilistic approach inspired by thermodynamics for calculating a function's global minimum. In our context, we search among a subset of potential rate profiles with simulated annealing to identify one with the lowest union bound value at a fixed SNR.

Algorithm \ref{algorithm1} summarizes the employed scheme. For a starting rate profile, we evaluate the value of the union bound at a fixed SNR as the cost of that rate profile; then, the subsequent rate profile is obtained by perturbation of a single bit-position. The perturbation is done by replacing a random bit-position from the information set with a random bit-location from the frozen set. We continue with the evaluation of the cost of the obtained rate profile and comparison with the cost of the previous one. If the cost of the new rate profile is less than the cost of the previous one, the new rate profile is kept. If the cost is higher, then the new rate profile is adopted with a certain probability depending on the difference of costs and the current temperature. If $E_k$ and $E_{k+1}$ are the costs of the $k$th and $(k+1)$th rate profiles, respectively, and $\Delta=E_{k+1}-E_k$, the probability of acceptance at temperature $T$ is given by
\begin{equation}
    p(\Delta)=\frac{1}{1+\exp\left(\frac{\Delta}{T}\right)} \approx \exp \left(\frac{-\Delta}{T}\right).
\end{equation}
In each step, when the algorithm accepts the $i$th new rate profile, temperature is set to $T_{max}\times a^{i}$, where $a$ is a parameter between 0 and 1, controlling the speed of the annealing. The acceptance probability is near one at the beginning of the algorithm when the temperature is high. This property enables the algorithm to accept some of the rate profiles that are not lower in the metric and continue with them. This step prevents the algorithm from being stuck at the local minima. This procedure continues until the $T_{min}$ is reached. 

We note that the search can be done either over all possible rate profiles or over the bit locations with the same RM score in non-full rank \cite{Moradi2021a} RM rate profiles to determine which of those bit locations should be added to the information set. It is worth mentioning that when the search is over the bit-locations with the same RM score, perturbation is done only among those that have the specific RM score in the frozen and the information sets. 

\section{Numerical Examples}

\begin{figure}[!t]
\centering
\includegraphics[width=3.8in,height=2in]{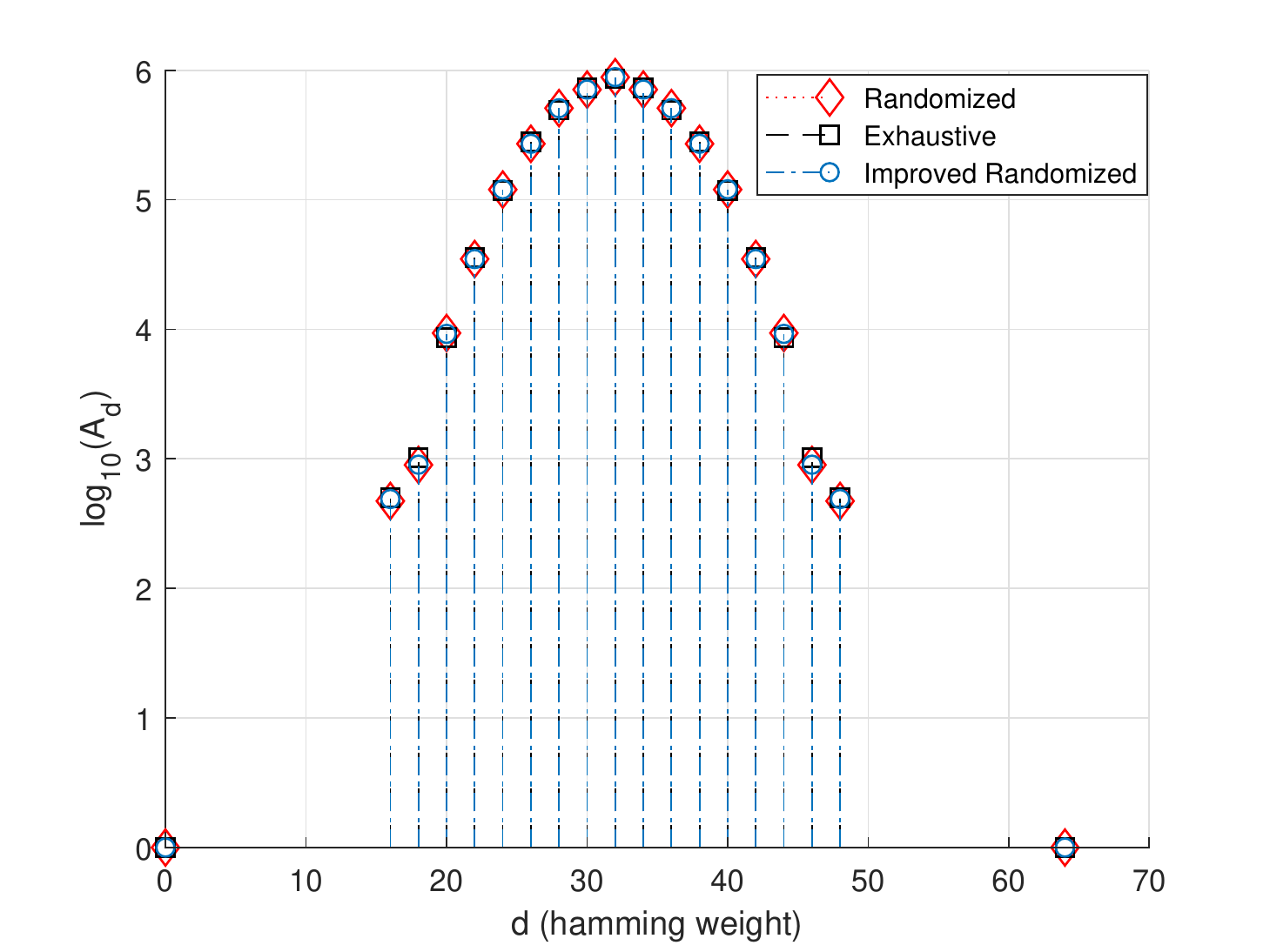}
\caption{Comparison of Approximated weight distribution of PAC (64,22) with RM rate profile and polynomial 133 with exact weight distribution generated by exhaustive search}
\label{fig1}
\end{figure}
      \begin{table}[t!]
\renewcommand{\arraystretch}{1.3}
\caption{Low-weight codewords of PAC (128,64) with polar and RM rate profiles}
\label{table1}
\centering
\resizebox{\columnwidth}{!}{%
\begin{tabular}{|r|c|c|c|c|c|}

\hline
\textbf{} & $A_8$ & $A_{12}$ & $A_{16}$ & $A_{18}$ & $A_{20}$\\
\hline
Polar profile in \cite{yao2021list} & 48 & 0 & 11032 & 6024 & $> 10^5$\\
\hline
Polar profile approx & 48 & 0 & 11274 & 2241 & 311486 \\
\hline
RM profile in \cite{yao2021list} & 0 & 0 & 3120 & 2696 & 95828\\
\hline
RM profile approx & 0 & 0 & 3285 & 1285 & 89563  \\
\hline
\end{tabular}
}
\end{table}
Let us now present several simulations and numerical results. In Figure \ref{fig1}, the weight distribution of a (64,22) PAC code with polynomial 133 in octal notation and the RM rate profile is approximated with two methods: 1) assuming $\boldsymbol{s}=0$ (labeled as ``Randomized"), and 2) assuming $\boldsymbol{s}\neq0$ (labeled as ``Improved Randomized"). The results are compared with the exact weight distribution obtained by an exhaustive search. The results show that both approximations give close results to the exact weight distribution in this scenario. In Table \ref{table1}, we compare the approximate weight distributions of PAC codes with codeword length 128 and rate $\frac{1}{2}$ with the low weight results obtained in \cite{yao2021list}, using the code polynomial 133 in octal notation with RM and polar rate-profiles. We observe that the approximate WEF results are very close to those obtained via exhaustive search.

\begin{figure}[t!]
    \centering
    \includegraphics[width=3.8in,height=2.36in]{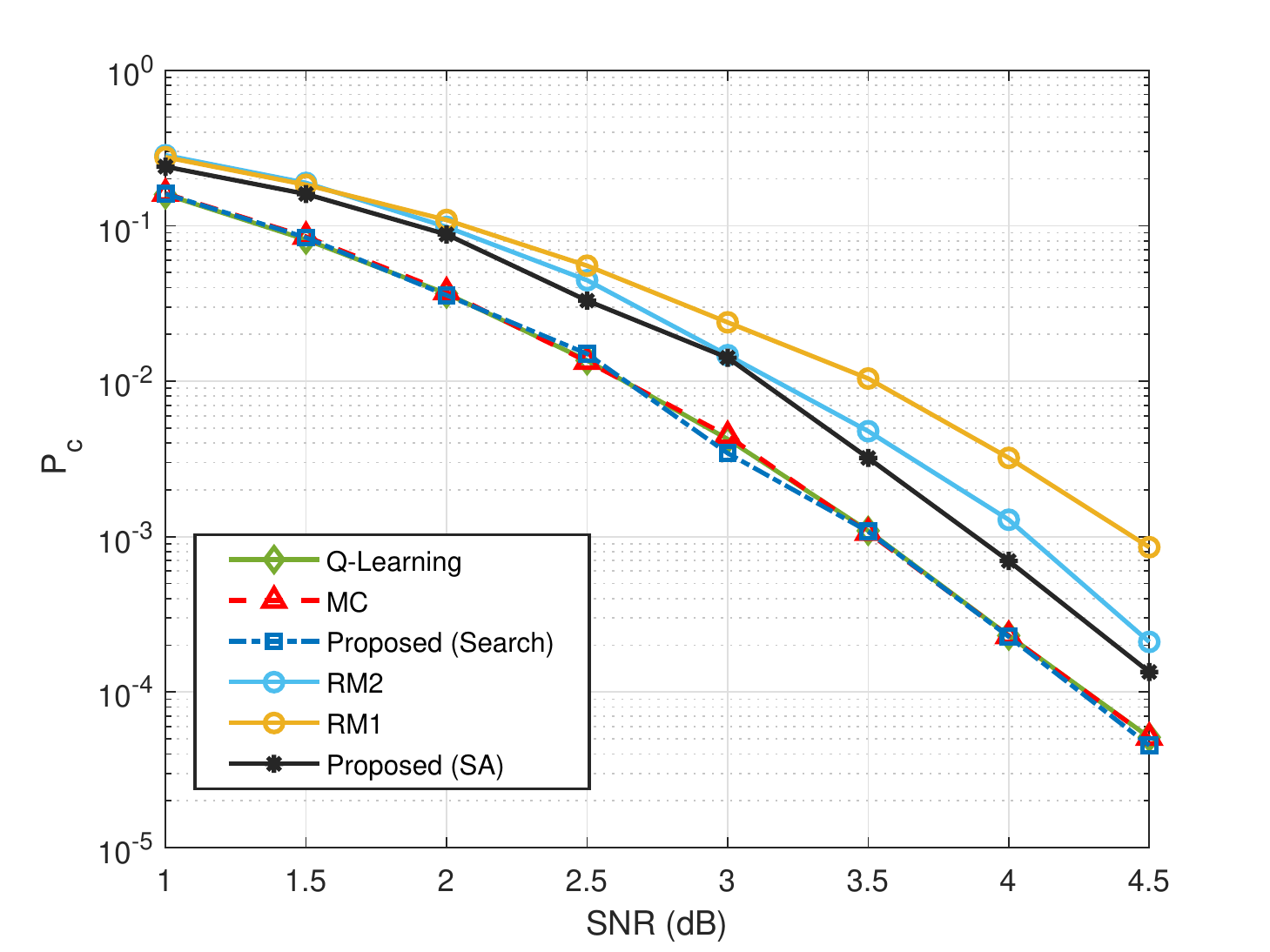}
    \caption{Performance of PAC (64,32) with polynomial 133 using different rate-profiles and list decoder (list size of 32)}
    \label{SAres1}
\end{figure}

In Fig. \ref{SAres1}, we design two rate profiles for a (64,32) PAC code, decoded with a list decoder and a list size of 32. One of the rate profiles is obtained by an exhaustive search over all possible combinations of bit-locations with an RM score of 3, along with bit-locations that have higher RM scores as information (non-frozen) part. The other is the output of the SA algorithm that searches among all possible rate profiles regardless of their RM score, obtained by setting $a=0.99$, $T_{max}=0.001$ and $T_{min}=10^{-4}$ with the SNR fixed at $3 dB$. The results show that the performance of the first one is similar to those of Monte-Carlo-based \cite{Moradi2021a} and Q-Learning \cite{mishra2021modified} designs. In this figure, $RM_1$ represents an RM rate profile that picks higher indexed bit-positions with the same RM score, while $RM_2$ picks lower indexed ones.

\begin{figure}[t!]
    \centering
    \includegraphics[width=3.8in,height=2.46in]{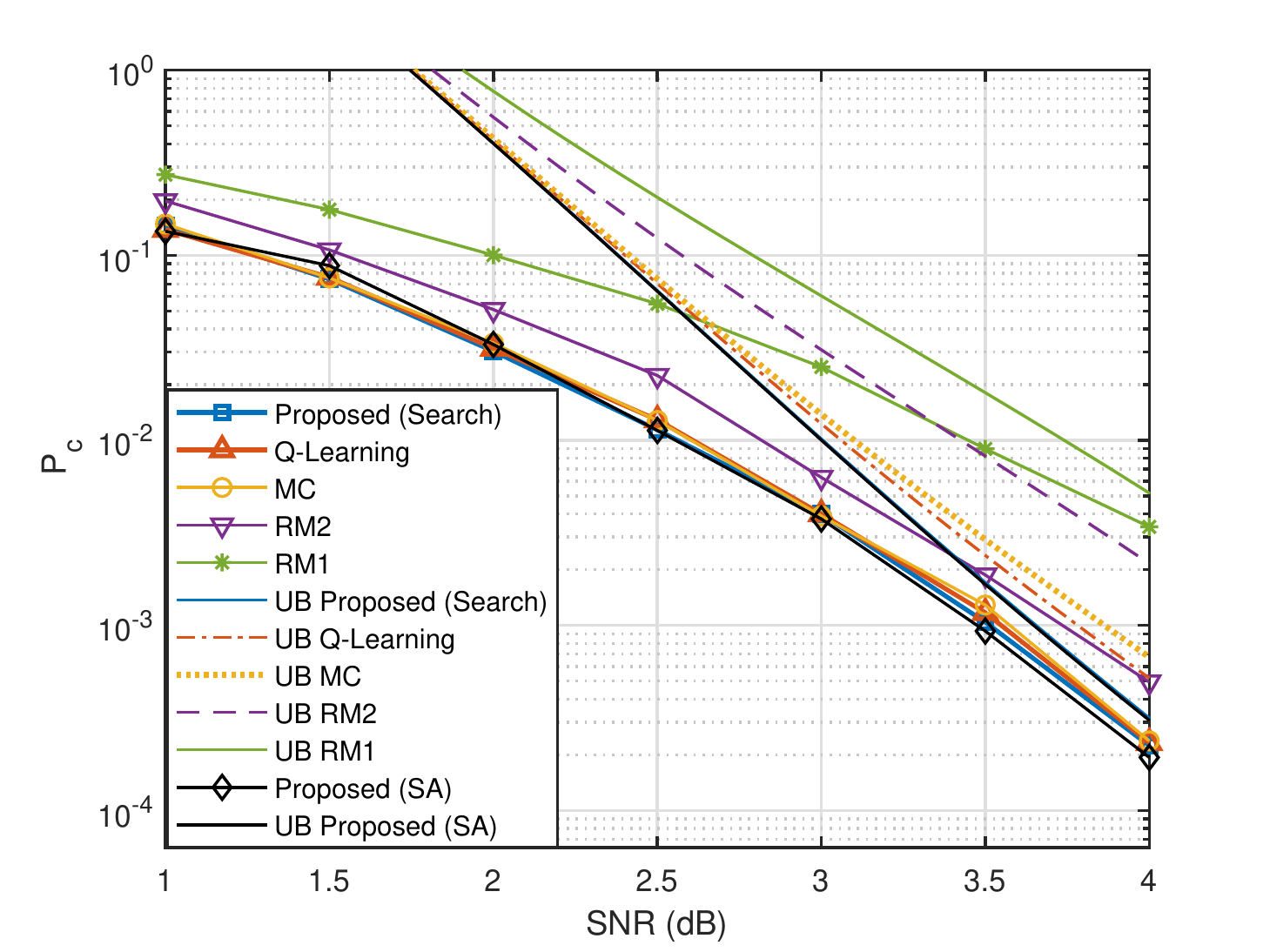}
    \caption{Performance of PAC (64,32) with different rate-profiles with list decoder (list size of 128) and polynomial 133 with their corresponding approximate union bounds}
    \label{SAres2}
\end{figure}
\begin{figure}[t!]
    \centering
    \includegraphics[width=3.8in,height=2.46in]{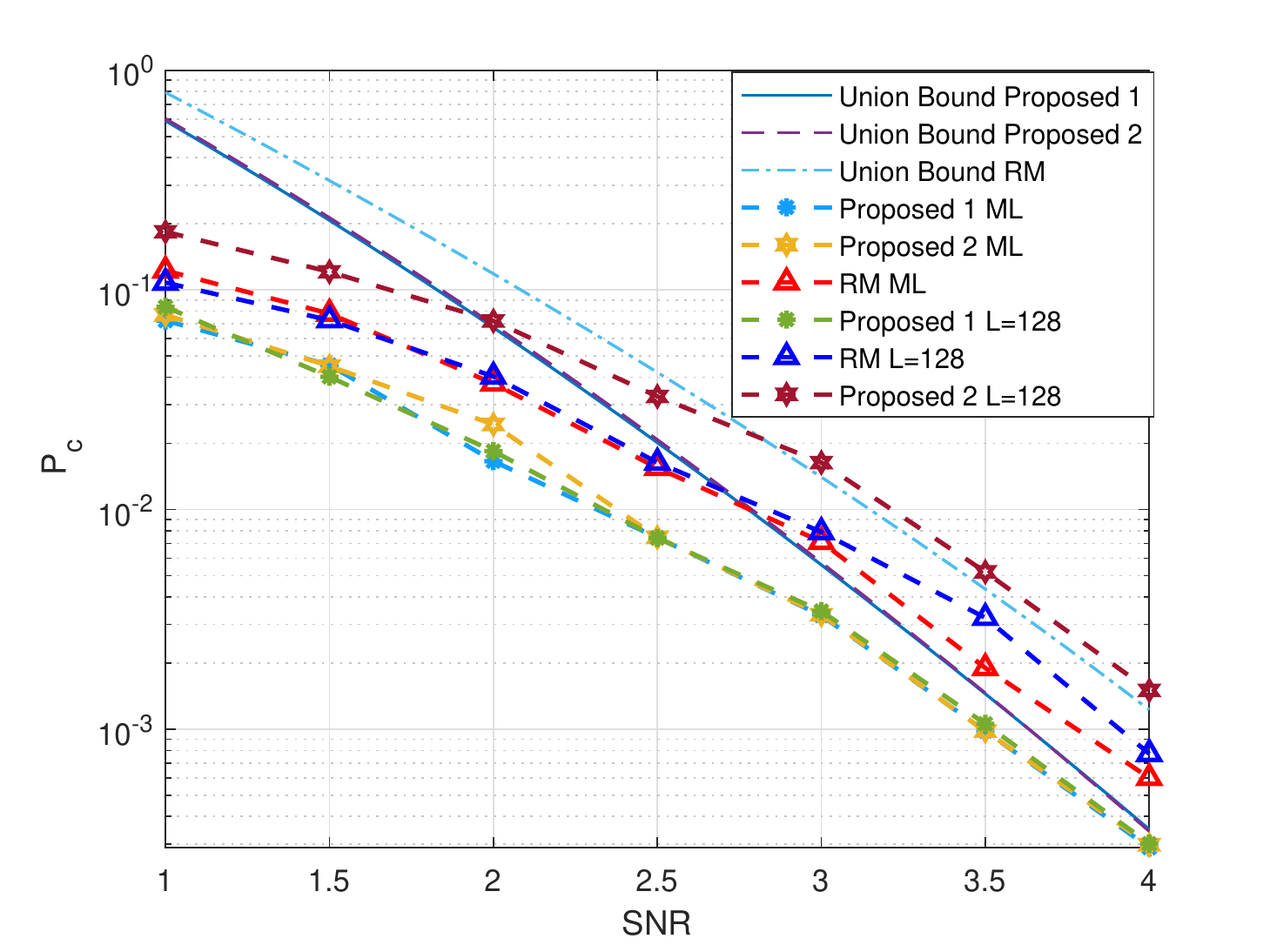}
    \caption{Performance of PAC(64,16) using polynomial 133 and different rate-profiles with ML decoder and List decoder (list size 128) and corresponding exact union bounds}
    \label{SAres3}
\end{figure}

 In Fig. \ref{SAres2}, the (64,32) PAC code with the same rate profiles as in Fig. \ref{SAres1} are decoded with a list decoder of size 128. The results show that their decoding performances match the corresponding union bounds (which assumes ML decoding). The SA-based design has an inferior performance with the list decoder with a list size of 32, but its performance is superior to the other rate profiles when the list size is 128.
In Fig. \ref{SAres3}, two rate profiles for a (64,16) PAC code are given. The codeword error rates with both ML decoder and list decoder with a list size of 128 are presented. As seen in Fig. \ref{SAres3}, one of the designed rate profiles (Proposed 2) has an inferior performance with the list decoder even with a large list size, while its performance with ML decoder matches the union bound. This shows that some rate profiles that are optimized under ML decoding may be inferior with suboptimal decoders. We attribute this behavior to the high mean computational complexity of the sequential decoding with the corresponding rate profiles. As noted in \cite{Moradi2021b}, since the convolutional encoded sequence in a PAC code sees polarized channels, the mean complexity of the sequential decoding is finite when $\ell R_{\ell} < \sum_{i=1}^{\ell} R_0 \left(1,W_{N}^{(i)}\right)$, for all $1<\ell<N$ where $N$ is the codeword length, $R_0 \left(1,W_{N}^{(i)}\right)$ is the cut-off rate of the $i$th polarized channel, i.e., $W_{N}^{(i)}$ and $R_{\ell}=\frac{\alpha}{\ell}$ with $\alpha$ being the number of non-frozen bit locations in the first $\ell$ bits of the rate profiled sequence (see \cite{Moradi2021b} for details). We remark that the PAC code with the rate profile 2 has a large mean computational complexity, and correspondingly, an inferior performance with the suboptimal decoders.
\label{sec5}

\section{Conclusions}

We have developed a probabilistic technique to compute the approximate weight distribution of PAC codes, which have an excellent match with the exact weight distributions that can be computed in a brute-force manner for small codes. We have also provided a method of using the computed weight distributions along with a union bound to design specific PAC codes, particularly, to select the code rate profiles using a discrete optimization method based on simulated annealing. Numerical examples demonstrate that the PAC codes with the designed rate profiles with the help of approximate union bounds offer high performance.
\label{conc}
\bibliographystyle{IEEEtran}
\bibliography{main}

\end{document}